# LLM Context Conditioning and PWP Prompting for Multimodal Validation of Chemical Formulas

Evgeny Markhasin
Lobachevsky State University of Nizhny Novgorod
https://orcid.org/0000-0002-7419-3605
https://linkedin.com/in/evgenymarkhasin

## Abstract

Identifying subtle technical errors within complex scientific and technical documents, especially those requiring multimodal interpretation (e.g., formulas in images), presents a significant hurdle for Large Language Models (LLMs) whose inherent error-correction tendencies can mask inaccuracies. This exploratory proof-of-concept (PoC) study investigates structured LLM context conditioning, informed by Persistent Workflow Prompting (PWP) principles, as a methodological strategy to modulate this LLM behavior at inference time. The approach is designed to enhance the reliability of readily available, general-purpose LLMs (specifically Gemini 2.5 Pro and ChatGPT Plus o3) for precise validation tasks, crucially relying only on their standard chat interfaces without API access or model modifications. To explore this methodology, we focused on validating chemical formulas within a single, complex test paper with known textual and image-based errors. Several prompting strategies were evaluated: while basic prompts proved unreliable, an approach adapting PWP structures to rigorously condition the LLM's analytical mindset appeared to improve textual error identification with both models. Notably, this method also guided Gemini 2.5 Pro to repeatedly identify a subtle image-based formula error previously overlooked during manual review, a task where ChatGPT Plus o3 failed in our tests. These preliminary findings highlight specific LLM operational modes that impede detail-oriented validation and suggest that PWP-informed context conditioning offers a promising and highly accessible technique for developing more robust LLM-driven analytical workflows, particularly for tasks requiring meticulous error detection in scientific and technical documents. Extensive validation beyond this limited PoC is necessary to ascertain broader applicability.

## 1. Introduction

The accurate validation of scientific and technical documents for subtle technical errors, particularly those requiring multimodal interpretation of elements like chemical formulas in images, remains a significant challenge. While Large Language Models (LLMs) are increasingly explored for complex analytical tasks, a fundamental aspect of their operation - robust pattern completion and intent recognition - often manifests as an inherent tendency to "correct" or make plausible inferences about perceived input imperfections. This behavior, though beneficial in many contexts, can inadvertently mask the very inaccuracies targeted during detail-oriented validation tasks. Addressing this challenge effectively, especially with readily available, general-purpose LLMs, necessitates methodological advancements for modulating these default behaviors at inference time. Such advancements ideally leverage domain expertise combined with advanced prompt engineering, offering a more accessible route than resource-intensive model modification or extensive fine-tuning for every specific validation nuance.

Building on a recent preprint [1] that proposed Persistent Workflow Prompting (PWP) as a methodology for such expert-driven, prompt-based guidance of general-purpose LLMs, and explored LLM context conditioning within a PWP-based proof-of-concept (PoC) prompt (*PeerReviewPrompt*) for critical manuscript review, this exploratory PoC study further investigates the application of these principles. Here, we focus on the specific challenge of validating chemical formulas within a single, complex test paper [2] known to contain both textual and image-based errors. This task, while involving a more defined analytical object (chemical formulas) than the "core methodology" focus of [1], requires full-document scope and presents a "needle in a haystack" problem, compounded by the LLM error-correction tendencies noted above.

This work is exploratory, based on observational assessment using the test paper [2]. The primary objectives were: (i) to conduct an initial evaluation of several prompting strategies (from simple direct prompts to more elaborate PWP-informed approaches incorporating task decomposition [3, 4] and self-reflection [5–7]) using generally available frontier reasoning models (Gemini 2.5 Pro and ChatGPT Plus o3 via chat interfaces); (ii) to perform an initial qualitative assessment of these models' multimodal analysis capabilities for this domain-specific task; (iii) to gauge whether basic prompting strategies showed sufficient promise, recognizing that poor performance on a challenging, yet circumscribed, test case could be a negative indicator; (iv) to identify a prompt design that could work reliably for this specific test case, rather than achieving optimized prompt text; and (v) to identify potential LLM behavioral issues, such as error suppression, and test potential solutions within the confines of this single test case. The PWP-informed *ChemicalFormulasValidationPrompt* developed herein, particularly its new task-specific core section, should be considered relatively unrefined.

This paper details these exploratory approaches and reports on their observed performance. We analyze LLM behaviors - such as apparent error suppression, inconsistent effort ("laziness"), and hallucinations - and discuss how PWP-informed context conditioning appeared to affect the reliability of chemical formula identification in our test case. Notably, observations with Gemini 2.5 Pro suggested the feasibility of multimodal error identification under these conditions (repeatedly identifying an image-based error overlooked by human review), an outcome not observed with ChatGPT Plus o3 in the same test. Ultimately, this study aims to highlight the potential of structured context conditioning as an accessible technique for adapting general-purpose LLMs for precise validation tasks in complex scientific documents, while clearly acknowledging its preliminary nature. To facilitate reproducibility and further investigation, links to sample AI analysis chats and the full Markdown-formatted text of the *ChemicalFormulasValidationPrompt* are provided in the Supporting Information.

## 2. Methodology

This study uses the same test publication [2] as in our prior work [1], known to contain demonstrable methodological flaws. The test paper file (also available via a link in Supporting Information) constitutes a combination of the main text and supporting information files (as available via paper's DOI [2]) totaling 44 pages. The same test paper presented a pertinent and, as it turned out, challenging test case for this task due to known, subtle errors in chemical formulas.

Specifically, page S-8 of the test paper's SI presents the formula for *ferrous ammonium sulfate* as $Fe(NH_4)_2SO_4$, which incorrectly omits one sulfate group (the correct formula for ferrous ammonium sulfate, Mohr's salt, is $(NH_4)_2Fe(SO_4)_2 \cdot 6H_2O$ or $(NH_4)_2Fe(SO_4)_2$ - anhydrous). The second known error is in the *hexamethyldisiloxane* formula shown on page 235 as spectral label in Figure 2(c), second from the bottom: $(CH_3)_3Si_2O$ (the correct formula is $((CH_3)_3Si)_2O$ or $(CH_3)_6Si_2O$).

While the first error is in text-based formula, the second error is in a raster image making this paper also suitable for initial tests of multimodal analysis. Considering, that the *test paper* contains 44 pages, this test case also presents a chemical "needle in a haystack" challenge.

All employed prompting techniques rely solely on standard prompt, requiring no API access, modification to the underlying models, or special tools. Further, all techniques employed target generally available models; specifically, Gemini Advanced 2.5 Pro and ChatGPT Plus o3 have been used in this study and accessed via the official chat bots (Gemini was also accessed via Google AI Studio). This deliberate restriction was adopted to maximize the potential for reproducibility of the observations presented. With these two models, we employed several prompting strategies, starting from basic direct prompts.

Our methodology involved (i) selecting a challenging test case with known errors, (ii) defining the specific errors to be targeted, (iii) systematically developing and testing a series of increasingly sophisticated prompting strategies with two leading LLMs, and (iv) analyzing the LLMs' behavior and performance, particularly focusing on the impact of context conditioning.

### a. Basic direct prompt

The naive approach asks the LLM directly to find errors: Find mistakes in chemical formulas and names. This prompt specifically mentions names, as names should generally be used to resolve formula errors.

### b. Decomposed prompt: extracted formula vs. extracted name

Because in chemistry communications, the general practice is that most formulas (except, perhaps, for the most basic ones) should have accompanying chemical names, a strategy was devised to direct the model to extract chemical names for each extracted formula and attempt identify problems by comparing (implicitly) the formulas with extracted names.

```
Execute the following task step-by-step:
1. Extract each and every chemical formula from the attached PDF.
2. For each extracted formula, extract every directly associated chemical name
   included in the text, if any.
3. For each extracted formula and associated names, consider if the chemical formula
   and associated names are correct and flag every formula/names combination that
   contains any errors.
4. For each flagged item, read the source PDF again and confirm that the item was
   extracted exactly. In case of any extraction errors, analyze the corrected item and
   consider if the flag should be removed.
5. Create a Markdown table that should include every flagged formula/names, clear
   description of any problems, corrected version, and clear reference location of the
   flagged items.
```

Here Step 4 attempts to elicit self-reflection to reduce errors and observed hallucinations.

### c. Decomposed prompt: extracted formula vs. generated formula

This prompt uses a different error detection workflow. Instead of comparing extracted formulas and names, it asks the LLM to generate names from extracted formulas. The idea here is that for minor errors, the LLM might be able generate correct name from incorrect formulas. The Step 3 instructs the LLM to generate a new formula from the previously generated name in the hope that this newly generated formula would be correct regardless of whether the extracted formula is correct. Finally, extracted and generated formulas are to be compared to identify potential errors.

```
Execute the following task step-by-step:
1. Extract chemical formulas of each and every chemical species containing at least
   two elements EXACTLY as they appear in the attached PDF.
2. For each extracted chemical formula generate associated name.
3. Convert each generated name to generated chemical formula.
4. For each generated chemical formula, determine if it matches previously extracted
   formula.
```

Here, the Step 1 attempts to reduce noise by filtering formula-like strings containing single or no chemical elements.

### d. PWP-based prompt with LLM context conditioning

Given the limitations of direct and simple generative approaches, a more robust strategy was explored, leveraging the context conditioning principles outlined in [1]. The *PeerReviewPrompt* prompt detailed in that work aimed to mitigate input bias through comprehensive context setting, and initial observations suggested this approach was effective in the test case presented therein. *ChemicalFormulasValidationPrompt* adapts core sections from the *PeerReviewPrompt*:

- Sections **I-III** (Core Objective, Persona, Context: Framework for Critical Review) were largely retained to establish the analytical mindset and operational guidelines.
- Section **V** (Final Instructions for Interaction) was kept to ensure consistent LLM behavior.
- Section **IV.A** (Foundational Principles & Workflow Application) was adapted.
- A new **Chemical Identifier Analysis** subsection was introduced into Section **IV** specifically for formula and name validation, providing a dedicated workflow for this task. Key instructions include: meticulous full-document scanning (text and figures) for chemical identifiers, initial attempt to define criteria for what constitutes a relevant chemical formula to reduce noise, a draft of explicit error categorization for formulas and names, protocols for multimodal analysis of figures, and a requirement for structured output detailing original entities, identified problems, proposed corrections, and exact source locations.

The full text of the prompt is available as a Markdown-formatted file from an OSF repository linked in Supporting Information and also as a PDF attachment.

## 3. Results and Discussion

### 3.1. Insights from Gemini Thinking Log

Initial exploratory tests using the basic direct prompt **2a** yielded inconsistent and generally unreliable results. While the text-based target error was occasionally identified, responses frequently included a significant amount of hallucinated errors. Interestingly, prominent types of hallucinations were quite specific, and plausible, matching the context set by the target text. LLMs specifically focused on oxygen symbol (capital letter O) replaced in chemical formulas with capital letter C (for carbon), digit zero, (occasionally, even uranium symbol) and a variety of falsely claimed subscripted oxygen symbols.

Both models also demonstrated "laziness". Sometimes they produced extensive output, riddled with hallucinated issues, sometimes they claimed having discovered no issues, and sometimes just a few candidate issues were reported. ChatGPT Plus o3 even demonstrated anthropomorphic complains in its partially exposed thought process, when it reasoned that "manually" going through the entire file searching for candidate formulas would take "forever" and that it needed to consider a different strategy. In other runs, ChatGPT could report that it had examined a part of the document or made a rough run, stating that specific requests for subsequent runs were necessary to discover additional candidate cases.

Decomposed prompts, such as those shown in **2b** and **2c** resulted only in marginal improvement: the text-based target error was identified somewhat more frequently (see sample runs [8, 9]), but the results remained unreliable and unsatisfactory.

Examination of Gemini's "Show thinking" logs (a feature providing insight into the model's processing steps) for various runs revealed a consistent pattern. With decomposed prompts, the LLM often correctly extracted the target formula (e.g., "$Fe(NH_4)_2SO_4$: Ferrous ammonium sulfate (Mohr's salt)"). However, in the subsequent validation step, it would sometimes erroneously mark the pair as correct, e.g.:

> Identified Chemical Formulas/Names and Initial Validation:
> ...(Other log lines)...
> $Fe(NH_4)_2SO_4$: Ferrous ammonium sulfate. Correct.

This observed behavior might stem from a core strength of LLMs: their inherent capability for error correction and understanding intent despite minor inaccuracies in the input. For instance, querying "What is the capital of grate britain?" typically yields "London", with the misspelling of "Great Britain" being automatically corrected. While usually beneficial, this default error tolerance becomes a hindrance when the objective is to detect such errors. This phenomenon is analogous to the "input bias" discussed in [1], where the LLM's tendency to accept input information as factual needs to be actively countered for critical evaluation.

In the context of formula validation, the LLM's natural inclination to reconcile a slightly incorrect formula with its correct accompanying name complicates direct "contrasting" methods (as in **2b**). An early attempt to address this problem was implemented as prompt **2c**, which aimed to avoid the direct contrasting of complementary entities, contrasting instead extracted and regenerated (and hopefully corrected) formulas instead. This approach, however, did not improve reliability of error detection. A more radical and systematic approach would be to affect the default "mindset" or mode of operation of LLMs through deliberate context conditioning.

### 3.2. LLM Context Conditioning

Context conditioning aims to selectively amend default tendencies of general purpose LLMs, matching the needs or nature of the task or group of tasks through a dedicated behavioral prompt section that should articulate desired LLM mode of operation, with particular emphasis on the targeted problematic aspects. One example of such a need is the critical treatment of input information to be analyzed in the course of critical manuscript analysis - the focus of the PWP-based *PeerReviewPrompt* [1]. During development of this prompt, an important behavioral aspect was identified that manifested as LLMs largely accepting input as factual information, not as object of scrutiny. While useful as a basis of in-context learning, this feature is at odds with critical analysis of a manuscript (manuscript must be provided to LLM as part of its input). Therefore, it was necessary to craft instructions that would selectively suppress "uncritical" behavior with respect to submitted manuscript.

In the case of the *PeerReviewPrompt*, the primary location for instructions countering uncritical behavior is Section **II**. Persona: Expert Critical Reviewer. While error correction tendency encountered in the present study is different from input bias, Section **II** of the prompt was used for the present task as well in the hope that the same critical attitude instilled in this section would also counter the error suppression feature.

Another aspect of LLMs behavior, as observed the in present study and discussed in the previous section, is their "laziness" by default. In general, if the prompt does not expressly state and emphasize the need for thorough analysis, the LLM will often do the least amount of work (or on the lower end), while being more thorough during other times, within the range consistent with possible prompt interpretations. This behavior was clearly observed, for example, in the case of the basic prompt **2a**, which asks the LLM to find mistakes and simply hopes that the model will go through the entire submitted document in detail. "Laziness" of LLM models manifests differently depending on model, temporary context (that is, the prior conversation history within the current conversation), and, possibly, even permanent context (if implemented and enabled), where models may also "remember" certain information from prior conversations. In fact, the ChatGPT o3 model occasionally expressed its "laziness" more or less clearly and even anthropomorphically in its responses and in its partially exposed thought process. Clearly, this kind of variation is not something one would want to see when performing critical analysis. Such variations would most likely guarantee incomplete analysis, overlooked "needles in a haystack", making the results unreliable and unusable.

This phenomenon manifested as laziness is, perhaps, of a more general nature. To be useful to wide audience and broad spectrum of tasks, general purpose LLMs need to be "creative" by producing varying or somewhat random output associated with each ambiguous, loosely constrained, or not constrained at all aspect within the range of possible prompt interpretations. In other words, if the prompt does not expressly states and emphasize the need for thorough analysis of the entire submitted document, LLM may and will "randomly choose" the scope and level of attention to details. To counteract this randomness, specific instructions on both scope and level of detail are incorporated in the *PeerReviewPrompt* Sections **I-III** (Core Objective, Persona, Context: Framework for Critical Review), with intentional repetition serving to emphasize their importance to the model.

There is yet another phenomenon to be discussed in this context, which is hallucinations. Extensive hallucinations were observed at the early stage of the present study involving basic prompts. Hallucinations may occur, for example, due to training knowledge gaps or due to faulty logic used for complex tasks [10–12]. Arguably, output errors (I intentionally try to use a more generic reference here), might also occur due to ambiguous instructions or insufficiently tight constraints included in the context. Instilling the various traits of a critical review in Section **II**, partial repetition in other sections as part of context conditioning, may also positively affect the level of hallucinations by providing additional constraints. Whether this hypothesis is correct or not needs to be investigated. However, extensive hallucinations observed at the early stage with simple prompts, have not been observed with the PWP-based prompt in this particular test case (though, some minor false positives were reported).

In case of the *PeerReviewPrompt*, Sections **I** (Core Objective) and **II** (Persona) are largely responsible for performing context conditioning. The *PeerReviewPrompt* prompt [1] designed for critical analysis of the core methodology (with the goal of subsequent development to extend the scope beyond the core methodology) in experimental chemistry manuscripts is largely relevant for the present task as well, as validation of chemical formulas and identifying any errors in them is an integral constituent of critical manuscript analysis. For this reason (and with the idea of subsequent integration of chemical formula validation workflow into *PeerReviewPrompt*), the "framing" sections of the original prompt were integrated into the formula-focused prompt mostly without modifications.

### 3.3. Advanced Validation and Multimodal Analysis

Observations from AI chats using the *ChemicalFormulasValidationPrompt* (Strategy **2d**) with Gemini 2.5 Pro [13] and ChatGPT Plus o3 [14] suggested potentially improved robustness in error identification compared to simpler prompts for this test case. Note that this prompt instructs the LLM to output a table of all extracted chemical formulas, detailing any identified issues and providing corrected versions or a check mark if no error is found. Crucially, in our tests with this PWP-based approach, both models consistently identified the target text-based error. Furthermore, due to the explicit instruction to perform multimodal analysis (specifically, analyzing figures), the Gemini 2.5 Pro model, across several trials, also identified the image-based error that had been missed in prior manual reviews. While occasional false positives were observed in the outputs from both models, the use of the PWP-based prompt with multimodal instructions appeared to result in a notable improvement in the detection of the known chemical formula errors in the test paper, including the formula embedded within the figure. In contrast, while the ChatGPT Plus o3 model is advertised as being capable of multimodal analysis, it failed to identify the error in the figure in this specific study.

### 3.4. Observed Differences in Gemini 2.5 Pro Performance Across Access Interfaces

An interesting observation during this study related to the performance of the Gemini 2.5 Pro model when accessed via different Google interfaces: the publicly available Gemini Advanced app (via gemini.google.com) and the developer-focused Google AI Studio. While both platforms theoretically provide access to the same underlying frontier model, and AI Studio offers extensive customization (though default settings were used in this work), qualitative differences in behavior were noted.

Our observational assessment, though not a systematic benchmark, suggested that the Gemini 2.5 Pro model accessed via AI Studio (with default parameters) often exhibited more consistent and precise behavior on the complex analytical tasks in this study compared to the version accessed via the Gemini Advanced app. This perceived enhanced performance manifested as potentially greater stability between runs, closer adherence to prompt instructions and user intent, and more accurate extraction of fine-grained details.

This difference was particularly evident in the limited multimodal analysis tests. While both interfaces enabled the model to identify the image-based formula error in the low-resolution figure within the test paper, the level of detail captured varied. Specifically, the Gemini 2.5 Pro model via the Gemini Advanced app repeatedly identified the formula in the figure as $(CH_3)_3SiO$, omitting the last subscript. In contrast, when accessed via AI Studio (defaults), the same nominal model repeatedly identified the defective formula more accurately as $(CH_3)_3Si_2O$, correctly including the last subscript. These specific observations, while based on a limited proof-of-concept, indicate that the access interface and its default configurations might influence an LLM's performance on nuanced, detail-oriented tasks. This observation highlights a practical consideration for researchers reporting or attempting to reproduce findings with nominally identical models accessed through different platforms.

### 3.5. Broader Implications, Limitations, and Future Directions

The observations from this proof-of-concept study, particularly regarding the apparent effectiveness of LLM context conditioning in managing certain LLM behaviors like error suppression (and input bias, as suggested in prior work [1]), point towards its potential utility beyond the specific task of chemical formula validation. While the findings herein are preliminary and derived from a limited testing scope, the principles of guiding LLM attention and operational mode through PWP-informed techniques could hold promise for broader applications. For instance, similar approaches might be valuable in fields such as medical AI, for workflows requiring meticulous processing and validation of information from patient records where precision is critical. Another relevant area could be the extraction and validation of data from semi-structured or poorly structured sources, common in pharmaceutical or technical documentation, where the ability to encourage an LLM to flag discrepancies rather than silently "correct" them may be highly desirable.

However, it is crucial to reiterate the limitations inherent in this exploratory work. The primary constraint is the reliance on a single test paper for evaluating the prompting strategies. Consequently, while the presented *ChemicalFormulasValidationPrompt* appeared effective in this specific context, these observations cannot be generalized without more extensive testing across diverse datasets and error types. The prompt itself, especially the "Chemical Identifier Analysis" workflow, remains an initial draft requiring further refinement. Future research should therefore prioritize rigorous testing of these PWP-informed context conditioning methods on a broader range of scientific documents to quantitatively assess their performance and generalizability. Such work should also include more systematic comparisons across different LLMs. Further refinement of the prompt architecture and a more controlled investigation into how specific conditioning instructions impact distinct LLM behaviors (e.g., error suppression, inconsistent effort, hallucinations) are also essential next steps to enhance the accuracy and ensure wider applicability of these techniques for complex scientific content analysis and validation.

## 4. Conclusions

This exploratory proof-of-concept (PoC) study investigated LLM-based validation of chemical formulas within a complex scientific document using a single test case with known errors. Observations indicated that simpler prompting strategies yielded unreliable results for the targeted errors, often compromised by LLM error-correction tendencies and inconsistent analytical effort, although they occasionally identified other untargeted issues like an imbalanced chemical equation, suggesting potential utility for broad exploratory testing. In contrast, a PWP-based approach with context conditioning appeared to improve the identification of the targeted error types.

Notably, despite the multimodal analysis instructions being largely adapted from prior work and not specifically refined for this task, the PWP-informed prompt guided Gemini 2.5 Pro to repeatedly identify a subtle error in an image-based chemical formula - an error previously overlooked by human review. This finding highlights the potential of systematically developed, context-conditioned prompts to uncover even untargeted or unexpected errors. These preliminary observations underscore the significant challenges LLMs face with detail-oriented validation tasks but also suggest that context conditioning may be a valuable technique for enhancing their reliability. The presented ChemicalFormulasValidationPrompt, though relatively unrefined, facilitated these initial qualitative assessments. Further research, beyond the scope of this limited PoC, is required to validate these findings and explore the full potential of such methods.

### Acknowledgments

Generative AI use has been an integral part of performed research, including interactive development of prompts via meta-prompting and extensive document revisions. This representative conversational log [15] documents the use of the Large Language Model Gemini (Google) to assist in the iterative revision and refinement of this manuscript. It serves as a demonstration of actively using AI as a peer collaborator during manuscript development. The documented interaction began with a draft manuscript that already included substantial preliminary revisions by the author.

### Supporting Information

**Note: the primary target model is Gemini Advanced 2.5 Pro.**

The prompt (*ChemicalFormulasValidationPrompt.md*) is included as a PDF attachment and is also available from: https://osf.io/nq68y/files/osfstorage?view_only=fe29ffe96a8340329f3ebd660faedd43.

To facilitate direct replication and review of the presented LLM analyses, the *test paper* (combined manuscript + SI [2]) PDF file used as input for the demonstrations is provided via a view-only link (Fair Use Statement): https://osf.io/nq68y/files/osfstorage?view_only=fe29ffe96a8340329f3ebd660faedd43.

Gemini 2.5 Pro (Google AI Studio) sample analysis chat - [13].

ChatGPT Plus o3 sample analysis chat - [14].

# A. Fair Use Statement - Sharing Test Paper

1. **Identification of Copyrighted Material:**

   - **Work:** "Enrichment of H217O from Tap Water, Characterization of the Enriched Water, and Properties of Several 17O-Labeled Compounds".
   - **Authors:** Brinda Prasad, Andrew R. Lewis, and Erika Plettner.
   - **Publication:** *Anal. Chem.* 2011, 83, 1, 231-239.
   - **DOI:** 10.1021/ac1022887.
   - **Publisher/Copyright Holder**: American Chemical Society.
   - **Material Shared:** A combined digital file containing the full text of the aforementioned article and its complete associated Supporting Information (SI).

2. **Sharing Mode:**

   - **Resource:** Private Open Science Framework (OSF) project repository.
   - **Location:** https://osf.io/nq68y/files/osfstorage?view_only=fe29ffe96a8340329f3ebd660faedd43.
   - **Protection Measures:** Due to private nature, the resource should not be indexed by search engines.

3. **Assertion of Fair Use:**

   The sharing of this copyrighted material is undertaken for specific, limited purposes, believed in good faith to constitute "fair use" under Section 107 of the U.S. Copyright Act (or applicable analogous principles in other jurisdictions).

4. **Purpose and Character of Use (Factor 1):**

   - **Non-Profit Educational and Research:** The use is strictly for non-commercial research and educational purposes, specifically within the context of scholarly critique and the advancement of research methodology.
   - **Transformative Use:** The work is not merely being reproduced; it is fundamentally repurposed as a research specimen for critical analysis. Its primary function in this context is not to convey its original purported findings, but to serve as the subject of rigorous evaluation and methodological demonstration.
   - **Critique and Commentary:** A core purpose is to conduct and disseminate a detailed, peer-review-like critique of the article's methodology, analysis, and conclusions. This critique identifies significant flaws within the original work.
   - **Advancement of Knowledge & Methodology:** The use includes the development and demonstration of a novel AI-driven prompt/technique for manuscript analysis. Sharing the specimen (the article + SI file) is integral to demonstrating and enabling the verification and further development of this new analytical method.

5. **Nature of the Copyrighted Work (Factor 2):**

   - The original work is a published scholarly article, typically factual in nature, a category often amenable to fair use for purposes of scholarship and critique.
   - However, the conducted analysis (central to this project) has revealed substantial flaws impacting the reliability and validity of the work's core research findings as presented. This impacts the assessment of its nature in the context of this specific use.

6. **Amount and Substantiality of the Portion Used (Factor 3):**

   The entire article and its complete Supporting Information are utilized and shared in a combined format.

   Justification: This amount is essential and necessary for the stated purpose. A comprehensive critique, akin to thorough peer review or forensic analysis, requires examination of the whole work, including all data and methods presented in the SI. Evaluating the integrity and validity of the research necessitates access to the complete context. Furthermore, the development and validation of the AI analysis prompt require the complete text as input. The combined file format, not available directly from the publisher, was the specific subject of the analysis.

7. **Effect of the Use upon the Potential Market for or Value of the Copyrighted Work (Factor 4):**

   - **No Harm to Legitimate Market:** This use is not intended to, nor is it likely to, negatively impact the legitimate market or value of the original copyrighted work. The publisher's market relies on the perceived value of the article as a source of valid scientific findings.
   - **Critique Reveals Lack of Value:** The critique resulting from this research demonstrates fundamental flaws undermining the article's claimed scientific value. Therefore, sharing the work specifically in this context (as a specimen for critique and methodological development) does not substitute for or usurp the market for the work based on its originally purported merits, as those merits are shown to be compromised. Dissemination for critique serves the public interest by highlighting these issues, distinct from fulfilling the original market demand.
   - **Controlled Access for Verification via Private OSF Project:** To ensure transparency and enable independent verification and follow-on research by interested parties engaging with the publicly disseminated research critique manuscript [TBD], the combined article + SI file (serving as the direct supporting evidence and test specimen) is hosted within a private Open Science Framework (OSF) project. A view-only link to this private project will be provided alongside the manuscript.
   - **Minimized Risk of Unintended Use:** This method ensures that access is granted specifically to individuals who are actively reviewing or assessing the research critique presented in the manuscript. The private nature of the OSF project prevents general public discovery through search engines, and the view-only restriction prevents facile downloading and redistribution. Access requires the specific link obtained from the context of the critique manuscript.
   - **Purpose Remains Transformative, Not Substitutive:** By utilizing a controlled-access, view-only repository linked directly to the research critique, this approach provides the necessary transparency for verification while strictly limiting potential downstream use and eliminating broad public access. This method strongly reinforces that the purpose is critique and verification (transformative uses), not market substitution for the original work's questioned scientific claims, thereby minimizing any potential harm to a legitimate market.

8. **Conclusion:**

   Based on the non-profit, educational, highly transformative nature of the use (critique, commentary, methodological advancement), the necessity of using the entire work for these specific purposes, and the argument that this use does not harm the legitimate market value due to the work's identified flaws and the distinct purpose of sharing, this distribution is asserted to be fair use.

   This material is intended solely for the recipient(s) for purposes directly related to verifying, understanding, or building upon the presented critique and methodological research. Further distribution is not permitted. Copyright remains with the original holder(s).